\documentstyle[12pt,epsfig]{article}
\newcommand{\myslash}[1]{#1 \llap /}
\newcommand{\partialslash}{\partial \llap /}
\newcommand{\VEV}[1]{\left\langle #1\right\rangle}
\newcommand{\mb}{\bar m}
\newcommand{\lsim}{
 \mathrel{\setbox0=\hbox{$<$}\raise0.6ex\copy0\kern-\wd0
 \lower0.65ex\hbox{$\sim$}}}
\newcommand{\gsim}{
 \mathrel{\setbox0=\hbox{$>$}\raise0.6ex\copy0\kern-\wd0
 \lower0.65ex\hbox{$\sim$}}}

\begin{document}

\thispagestyle{empty}
\begin{flushright}
       ADP-96-25/224, TUM/T39-96-22 \\
\end{flushright}
\vspace*{1cm}

\begin{center}
{\large\bf Deep Inelastic Structure Functions in a 
           Covariant Spectator Model} \\
\vspace{1cm}
K.\ Kusaka$^{a\dagger}$, G.\ Piller$^{b}$, A.W.\ Thomas$^{c,d}$ and 
A.G.\ Williams$^{c,d}$\\

\vspace*{1cm}

{\it
$^a$Department of Physics, Tokyo Metropolitan University, Tokyo 192, Japan
} \\
\vspace{.2cm}
{\it
$^b$
Physik Department, Technische Universit\"at M\"unchen, 
D-85747 Garching, Germany
} \\
\vspace{.2cm}
{\it
$^c$Department of Physics and Mathematical Physics, University of Adelaide,\\
S.A. 5005, Australia
} \\
{\it
$^d$Institute for Theoretical Physics, University of Adelaide, 
S.A. 5005, Australia
} \\

\vspace*{3cm}

\begin{abstract}

\noindent
Deep-inelastic structure functions are studied within a covariant scalar 
diquark spectator model of the nucleon.  
Treating the target as a two-body bound state 
of a quark and a scalar diquark, the Bethe-Salpeter equation (BSE) for 
the bound state vertex function  is solved in the ladder approximation.  
The valence quark distribution is discussed in terms of the solutions of 
the BSE.

\vspace{1cm}
\end{abstract}

\end{center}
\vfill
${}^\dagger$ JSPS Research Fellow
\begin{flushleft}
\begin{tabbing}
E-mail: \={\it kkusaka@phys.metro-u.ac.jp}\\
\>{\it gpiller@physik.tu-muenchen.de}\\
\>{\it athomas, awilliam@physics.adelaide.edu.au}
\end{tabbing}
\end{flushleft}
\newpage

\section{Introduction}

In recent years many attempts have been made to understand 
the nucleon structure functions measured in lepton deep-inelastic 
scattering (DIS).  
Although perturbative QCD is successful in describing the 
variation of structure functions with the squared momentum transfer, 
their magnitude and shape is governed by the non-perturbative physics of 
composite particles, and is, so far, not calculable directly from QCD. 

A variety of models has been invoked  to describe    
nucleon structure functions. 
Bag model calculations for example, 
which are driven by the dynamics of quarks bound 
in a nucleon bag,
quite successfully describe  non-singlet unpolarized and polarized 
structure functions (see e.g. \cite{Jaffe75,ScThLo89} and references 
therein).
However such calculations are not relativistically covariant.

A covariant approach to nucleon structure functions is given by 
so called ``spectator models'' \cite{MeyMul91,MeScTh94,MePiTh95}.
Here the leading twist, non-singlet quark distributions are 
calculated from the process in which the target nucleon splits into 
a  valence quark, which is scattered by the virtual photon, 
and a  spectator system carrying baryon number $2/3$.  
Furthermore the spectrum of spectator states is assumed to be 
saturated through single scalar and vector diquarks.  Thus, 
the main ingredient of these  models are covariant quark-diquark 
vertex functions.  

Until now vertex functions have been merely parameterized such that the 
measured quark distributions are reproduced, and no attempts have been made 
to connect them to some dynamical models of the nucleon.  
In this work we construct the vertex functions from a model Lagrangian 
by solving the Bethe--Salpeter equation (BSE) for the quark-diquark 
system.  
However, we do not aim 
at a detailed, quantitative description of nucleon structure functions 
in the present work.  
Rather we outline how to extract quark-diquark vertex functions from 
Euclidean solutions of the BSE. 
In this context several simplifications are made.  
We consider only  scalar diquarks as spectators and restrict ourselves 
to the $SU(2)$ flavor group.  
The inclusion of pseudo-vector diquarks and the generalization to 
$SU(3)$ flavor are relatively straightforward extensions 
and will be left for future work. 
It should be mentioned that the quark-diquark Lagrangian used here 
does not account for quark confinement inside nucleons. 
However the use of a confining quark-diquark interaction  
should also be possible within the scheme that we use.

As an important result of our work
we find that the vertex function of the nucleon is 
highly relativistic even in the case of weak binding. 
Furthermore we observe that the 
nucleon structure function, $F_1$, is determined to a large extent by 
the relativistic kinematics of the quark-diquark system and is 
not very sensitive to its dynamics as long as the spectator system is 
treated as a single particle.

The outline of the paper is as follows. In Sec.\ref{spectatorModel}
we introduce the spectator model for deep-inelastic scattering. 
Section \ref{diquarkModel} focuses on the scalar diquark model 
for the nucleon which yields the quark-diquark vertex function as a 
solution of a ladder BSE.
In Sec.\ref{numerical} we present numerical results for the 
quark-diquark vertex function and the nucleon structure function, 
$F_1$. 
Finally we summarize and conclude in Sec.\ref{summary}.

\section{Deep-Inelastic Lepton Scattering in the 
\protect \\Spectator Model}
\label{spectatorModel}

Inclusive deep-inelastic scattering of leptons from hadrons is described 
by the hadronic tensor 
\begin{eqnarray}
W^{\mu \nu}(P,q)
&=& \frac{1}{2\pi} \int d^4\xi\, e^{iq\cdot \xi}
\left\langle P\left|J^{\mu}(\xi)\,  J^{\nu}(0)
                 \right|P
\right\rangle,
\end{eqnarray}
where $P$ and $q$ are the four-momenta of the target and exchanged, 
virtual photon respectively, and $J^{\mu}$ is the hadronic
electromagnetic current.
In unpolarized scattering processes only the symmetric piece of 
$W^{\mu \nu} = W^{\nu \mu}$ is probed. 
It can be expressed in terms of two structure functions $F_1$ 
and $F_2$, which depend on the Bjorken scaling variable, $x = Q^2/2 P\cdot q$, 
and the squared momentum transfer, $Q^2 = -q^2$: 
\begin{equation}
W^{\mu \nu}(q,P) = 
                \left( -g^{\mu\nu}+\frac{q^\mu q^\nu}{q^2} \right) 
                         F_1(x,Q^2) +
                \left( P^\mu - q^\mu \frac{P\cdot q}{q^2} \right)
                        \left( P^\nu - q^\nu \frac{P\cdot q}{q^2} \right)
                        \frac{F_2(x,Q^2)}{P\cdot q}. 
\end{equation}
In the Bjorken limit ($Q^2, P\cdot q \rightarrow \infty$; but finite $x$) 
in which we work throughout, 
both structure functions depend up to logarithmic corrections on $x$ only,  
and are related via the Callan-Gross relation: 
$F_2 = 2 x F_1$. 

The hadronic tensor, $W^{\mu \nu}$, is connected via the optical theorem to
the amplitude $T^{\mu \nu}$ for virtual photon-nucleon forward 
Compton scattering: 
\begin{equation}
        \frac{1}{\pi}\hbox{Im}\,T^{\mu\nu}(q,P) =  W^{\mu\nu}(q,P).
        \label{dispersion}
\end{equation}
In the Bjorken limit the interaction of the virtual photon with  a  
valence quark from the target leads to a spectator system carrying 
diquark quantum numbers, i.e. baryon number $2/3$ and spin $0$ or $1$.  
In the spectator model it is assumed that the spectrum of spectator 
states can be saturated  through a single scalar and pseudo-vector 
diquark \cite{MeyMul91,MeScTh94,MePiTh95}. 
In the following we will restrict ourselves to contributions from 
scalar diquarks only. The generalization to include a pseudo-vector
diquark contribution is left for future work.
The corresponding Compton amplitude is 
(Fig. \ref{fig:diquarkSpectatorForT}):
\begin{eqnarray}
T^{\mu\nu}_S(q,P) & = &
\VEV{\frac{5}{6}+\frac{\tau_3}{2}}_N
                \int \frac{d^4k}{(2\pi)^4 i} \bar u(P)\bar\Gamma(k,P-k)
                        S(k)\gamma^\mu S(k+q)\gamma^\nu S(k)
        \label{compton}\\
        & & \qquad\qquad\qquad\qquad\qquad\qquad
                \times D(P-k) \Gamma(k,P-k) u(P),
        \nonumber 
\end{eqnarray}
where the flavor matrix has to be evaluated in the nucleon iso-spin space.  
The integration runs over the quark momentum $k$. 
The Dirac spinor  of the spin averaged nucleon target with momentum $P$ 
is  denoted by $u(P)$.  
Furthermore $S(k)=1/(m_q-\myslash{k} - i\epsilon)$ and 
$D(k)=1/(m_D^2-k^2-i\epsilon)$ are the propagators of the quark 
and diquark respectively, while $\Gamma$ is the quark-diquark vertex 
function.  To obtain the hadronic tensor,
the scattered quark and the diquark spectator have to be put on mass-shell
according to Eq.(\ref{dispersion}):  
\begin{eqnarray}
        S(k+q) & \rightarrow & i\pi \delta(m_q^2-(k+q)^2)
                                (m_q+\myslash{k}+\myslash{q}),
        \nonumber \\
        D(P-k) & \rightarrow & i\pi \delta(m_D^2-(P-k)^2).
\label{onshell}
\end{eqnarray} 
The vertex function $\Gamma$ for the target, which  in our 
approach is a positive energy, spin-1/2 composite state of a quark and a 
scalar diquark, is given by two independent Dirac structures:
\begin{equation}
        \left. \Gamma(k, P-k)\right|_{(P-k)^2=m_D^2} = 
                \left( f^{\rm on}_1(k^2) + 
                        \frac{2\myslash{k}}{M}
                                f^{\rm on}_2(k^2) 
                \right)\Lambda^{(+)}(P), 
        \label{DISpara}
\end{equation}
where $\Lambda^{(+)}(P) = 1/2 + \myslash{P}/2M$ is the 
projector onto positive energy spin-1/2 states and 
$M=\sqrt{P^2}$ is the invariant mass of the nucleon target.  
Note that according to the on-shell condition in Eq.(\ref{onshell}) 
the scalar functions, $f^{\rm on}_{1/2}$, will depend on 
$k^2$ only.

{}From Eqs.(\ref{dispersion} -- \ref{DISpara}) we then obtain 
for the valence quark contribution to the structure function $F_1$: 
\begin{eqnarray}
        F_{1}^{\rm val.}(x)&=&
\VEV{\frac{5}{6}+\frac{\tau_3}{2}}_N
                \frac{1}{16\pi^3}\int_{-\infty}^{k^2_{\rm max}}\,
                        \frac{d k^2}{m_q^2 - k^2}
        \nonumber\\
        & &\times\left\{ \left( 1 - x + 
                        \frac{( m_q + M )^2 - m_D^2}{ m_q^2 - k^2 } x 
                \right) \frac{f^{\rm on}_1(k^2)^2}{4}
                \right.
        \label{FofX}\\
        & &\qquad
                 - \left( 1+x + \frac{2\,m_q}{M} x + 
                   \left( 1 - \frac{2\,m_q}{M} \right) \,
                       \frac{ ( m_q + M )^2 - m_D^2 }
                                { m_q^2 - k^2 } x 
                \right) \frac{f^{\rm on}_1(k^2)\,f^{\rm on}_2(k^2)}{2}
        \nonumber\\
        & &\qquad 
                + \left( 
                4 \frac{m_q^2-k^2}{M^2}
                + \left(1 - \left({2m_q\over M}\right)^2\right)(1+2x)
                \right.
        \nonumber\\
        & &\qquad\qquad\left.\left.
                + \left(1-\frac{2\,m_q}{M} \right)^2\,x
                + \left(1-\frac{2\,m_q}{M} \right)^2\,
                        \frac{ ( m_q + M )^2 - m_D^2 }{m_q^2-k^2}\,x
                \right) \frac{f^{\rm on}_2(k^2)^2}{4}
        \right\}.
        \nonumber
\end{eqnarray}
The upper limit of the $k^2$--integral is denoted by:
\begin{equation}
        k^2_{\rm max}=x\left( M^2 -\frac{m_D^2}{1-x}\right),   
        \label{k2max}
\end{equation}
Note that $k^2_{\rm max} \rightarrow -\infty$ for 
$x\rightarrow 1$. This implies that for any 
regular vertex function 
$F_{1}^{\rm val.} \rightarrow 0 $ for $x\rightarrow 1$ and thus the
structure function automatically has the correct support.

Since the spectator model of the nucleon is valence-quark dominated,
the structure function $F_1^{\rm val.}$ in Eq.(\ref{FofX}) is  
identified with the leading twist part of $F_1$ at some typical
low momentum scale, $\mu^2 \lsim 1$ GeV$^2$. 
The physical structure function at large 
$Q^2 \gg \mu ^2$ is then to be generated via $Q^2$-evolution.

It should be mentioned that the Compton amplitude in Eq.(\ref{compton}) 
and also the expression for the structure function in (\ref{FofX}) 
contain poles from the quark propagators attached to the 
quark-diquark vertex functions.
{}From Eq.(\ref{k2max}) it follows that these 
poles do not contribute when $M < m_D + m_q$.
This condition is automatically satisfied
if the nucleon is considered as a bound state of the quark and 
diquark, as done in the following.

In the next section we shall determine the vertex function $\Gamma$, 
or equivalently  $f_1^{\rm on}$ and $f_2^{\rm on}$ from Eq.(\ref{DISpara}) 
as  solutions of a ladder BSE.

\section{Scalar Diquark Model for the Nucleon}\label{diquarkModel}

We now determine the vertex function (\ref{DISpara}) 
as the solution of a BSE for a quark-diquark system. 
We start from the following Lagrangian:
\begin{eqnarray}
        &{\cal L} = & \bar\psi_a\left(i \partialslash - m_q\right)\psi_a
                + \partial_\mu\phi_a^*\partial^\mu\phi_a - m_D^2 \phi_a^*\phi_a
        \label{action}\\
        & & + i \frac{g}{2\sqrt{2}} \epsilon^a_{bc}
                \psi^T_b C^{-1}\gamma_5 \tau_2\psi_c \, \phi^*_a
                -i \frac{g}{2\sqrt{2}} \epsilon^a_{bc}
                \bar\psi_b C^{-1}\gamma_5 \tau_2\bar\psi^T_c \, \phi_a, 
\nonumber
\end{eqnarray}
where we have explicitly indicated $SU(3)$ color indices but 
have omitted flavor indices.
We restrict  ourselves to flavor  $SU(2)$, 
where $\tau_2$ is the symmetric generator which acts  on the 
iso--doublet quark field, $\psi$, with mass $m_q$.  
The charged scalar field, $\phi$, represents  the flavor--singlet
scalar diquark carrying an invariant  mass $m_D$.  
Similar Lagrangians have been used recently to 
describe some static properties of the nucleon, such as its mass 
and electromagnetic charge (see e.g. \cite{BuAlRe92,HuaTjo94,AsIsBeYa95}).

The nucleon with 4--momentum $P$ and spin $S$ 
is described by the bound state Bethe-Salpeter (BS) vertex function $\Gamma$:  
\begin{equation}
        \hbox{F.T.} \left\langle 0\left| 
                        T \psi_a(x)\phi_b(y) 
                \right|P,S\right\rangle
        = \delta_{ab} S(k) D(P-k) i \Gamma(k,P-k) u(P,S).
\end{equation}
Here $u(P,S)$ is the nucleon Dirac spinor and 
F.T. stands for the Fourier transformation.~\footnote{
We use the normalization 
$ \VEV{P^\prime | P}= 2 P^0(2\pi)^3 \delta^{(3)}(\vec P^\prime-\vec P)$ 
and $\sum_{S} u(P,S)\bar u(P,S)= \sqrt{P^2} + \myslash{P}= M+\myslash{P}$.}
(Again we have omitted  $SU(2)$ flavor  indices.)

We will now discuss the integral equation for the vertex function 
$\Gamma$ in the framework of the ladder approximation.

\subsection{Ladder BSE}

For the following discussion of the integral equation for the vertex function 
$\Gamma$ we write the quark momentum as $q + \eta_q P$ and the diquark 
momentum as $-q + \eta_D P$. 
The weight factors $\eta_q$ and $\eta_D$ are 
arbitrary constants between 0 and 1 and satisfy $\eta_q + \eta_D = 1$.  
Within the ladder approximation the BSE for the vertex function 
of  a positive energy, spin-1/2 model nucleon can be written as 
(see Fig.(\ref{fig:ladderBSE})):  
\begin{equation}
        \Gamma(q,P)u(P,S)=g^2 \int\frac{d^4k}{(2\pi)^4 i}
                S(-k-q-(\eta_q-\eta_D)P) S(\eta_q P+k) D(\eta_D P-k) 
                \Gamma(k,P)u(P,S),
        \label{ladderBSE}
\end{equation}
where the flavor and color factors have  already been worked out.  
The scattering kernel is given by a $u$--channel quark exchange 
according  to the interaction Lagrangian in Eq.(\ref{action}). 

Since we are only interested in positive energy solutions, 
we may write the vertex function as:   
\begin{equation}
        \Gamma(q,P)=\left( a f_1(q,P) + b f_2(q,P)
        + \frac{\myslash{q}}{M} f_2(q,P) \right)
                \Lambda^{(+)}(P).
        \label{BSvertexPara}
\end{equation}
The arguments of the scalar functions $f_\alpha(q,P)$ 
are actually $q^2$ and $P\cdot q$, but we use this shorthand
notation for brevity.
With $a$ and $b$ we denote as yet unspecified scalar functions of 
$q^2$ and $P\cdot q$ which will be chosen later for convenience.  
(The definition of $f_{1/2}^{\rm on}$ in Eq.(\ref{DISpara}) corresponds 
to a specific choice of $a$ and $b$.)

\subsection{Wick Rotation}\label{wickRotation}

After multiplying the BSE in (\ref{ladderBSE}) with appropriate 
projectors (which depend on $a$ and $b$), we obtain a pair of coupled
integral equations for the scalar functions $f_1(q,P)$ and $f_2(q,P)$:  
\begin{eqnarray} \label{prjctd_BSE}
        &f_\alpha(q,P)=g^2 \int \frac{d^4k}{(2\pi)^4i}\,
        & \tilde D_q(-q-k-(\eta_q-\eta_D)P)\\   
        & &\times D_q(\eta_qP+k)\,D_D(\eta_qP-k)\,
        K_{\alpha\beta}(q,k,P)\, f_\beta(k,P), \nonumber
\end{eqnarray}
where $D_q(p)\equiv 1/(m_q^2-p^2-i\epsilon)$ and
$D_D(p)\equiv 1/(m_D^2-p^2-i\epsilon)$ 
are the denominators of the quark and diquark propagators, respectively.  
The indices $\alpha$ and $\beta$ stand for the independent Dirac structures
of the vertex function $\Gamma$, 
i.e. in the scalar--diquark model they run from 1 to 2 
according to (\ref{BSvertexPara}).  
Consequently the function 
$K_{\alpha \beta}(q,k,P)$ is  a $2\times 2$ matrix, 
where its explicit form depends on the 
definition of the scalar functions $f_\alpha(q,P)$.   
We use a form factor for the quark--diquark coupling which 
weakens the short range interaction between the quark and the diquark and 
ensures the existence of solutions with a positive norm.  
For simplicity, we use a $u$--channel form factor which can be
conveniently absorbed into 
the denominator of the exchanged quark propagator as follows:
\begin{equation}
        D_q(p) \rightarrow \tilde D_q(p) \equiv 
        D_q(p) \frac{\Lambda^2}{\Lambda^2 - p^2-i\epsilon}.
        \label{exchD}
\end{equation}
As a next step let us 
analyze the singularities of the integrand in  Eq.(\ref{prjctd_BSE}).  
For this purpose we choose the nucleon rest frame 
where $P_\mu=P^{(0)}_\mu\equiv(M,\vec 0\,)$ 
and put the weight constants $\eta_q$ and $\eta_D$ to the classical values:
\begin{eqnarray}
        \eta_q & \equiv & \frac{m_q}{m_q+m_D} \equiv \frac{1-\eta}{2}\;,
        \label{etaq} \\
        \eta_D & \equiv & \frac{m_D}{m_q+m_D} \equiv \frac{1+\eta}{2}\;.
        \label{etad}
\end{eqnarray}
Here we have introduced the asymmetry  parameter 
$\eta\equiv\frac{m_D-m_q}{m_q+m_D}$, such that the invariant quark and 
diquark mass is  given by  $m_q=\mb(1-\eta)$ and $m_D=\mb(1+\eta)$ 
respectively, where  $\mb=(m_q+m_D)/2$.  
In the complex $k_0$ plane $D_q(\eta_qP+k)$ and $D_D(\eta_qP-k)$ 
will be singular for:
\begin{eqnarray}
         & k^0 & = -\eta_q  M \pm E_q(\vec k) \mp i\epsilon,
        \label{qcut} \\
         & k^0 & = \eta_D  M \pm E_D(\vec k) \mp i\epsilon,
        \label{dcut} 
\end{eqnarray}
where $E_q(\vec k)=\sqrt{m_q^2+{\vec k\,}^2}$ and 
$E_D(\vec k)=\sqrt{m_D^2+ {\vec k\,}^2}$.  
The cuts lie in the second and forth quadrant of the complex 
$k_0$-plane. 
However for a bound state,  $0 < M < m_q+m_D$, a gap occurs between these 
two cuts which includes  the imaginary $k_0$ axis.

Next, consider the singularities of the exchanged quark propagator:  
\begin{equation}
        k^0 = -q^0 +\eta M \pm E_{m_i}(\vec q+\vec k) \mp i\epsilon,
        \label{exch_pole}
\end{equation}
where $E_{m_i}(\vec k)=\sqrt{m_i^2+{\vec k\,}^2}$ and $m_i = m_q, \Lambda$ 
for $i=1, 2$, respectively.
The sum of the second and third term at the RHS of 
Eq.(\ref{exch_pole}) is bound by:
\begin{eqnarray}
         \eta M + E_{m_i}(\vec q+\vec k) & \ge  (m_D-m_q) 
                           \frac{M}{m_q+m_D} + m_i,
        \label{bound1} \\
        \eta M - E_{m_i}(\vec q+\vec k) & \le  (m_D-m_q) 
                            \frac{M}{m_q+m_D} - m_i.
        \label{bound2}
\end{eqnarray}
The diquark should be considered as a bound state of 
two quarks which implies $m_D < 2 m_q$.
Together with $m_i \geq m_q$, namely setting the form factor mass 
$\Lambda$ larger than $m_q$, 
we have $m_D-m_q+m_i>0$ and $m_D-m_q-m_i<0$. 
Consequently we find from Eqs.(\ref{bound1},\ref{bound2}) 
$\eta M + E_{m_i}(\vec q+\vec k) >0$ and 
$\eta M - E_{m_i}(\vec q+\vec k) < 0$ for any momenta 
$\vec q$ and $\vec k$.  
Therefore,  if $-q^0 +\eta M - E_{m_i}(\vec q+\vec k) > 0$ or 
$-q^0 +\eta M + E_{m_i}(\vec q+\vec k) <0$ a so-called ``displaced''
pole will occur in the first or  third quadrant, respectively.  
In other words, the displaced-pole-free condition is:
\begin{equation}
        \eta M - E_{m_i}(\vec q+\vec k) < q^0 < 
        \eta M + E_{m_i}(\vec q+\vec k),
        \label{cndition0}
\end{equation}
for any $\vec k$.  Since $\vec k$ is an  integration variable, 
$E_{m_i}(\vec q+\vec k)$ will adopt its minimum value, $m_q$, at  
$\vec k = -\vec q$ for $i=1$. The above condition therefore simplifies to:  
\begin{equation}
        (q^0-\eta M)^2 < m_q^2.
        \label{condition_halfWR}
\end{equation}
If $q^0$ is Wick rotated to pure imaginary values, i.e.  
$q^{\mu} \rightarrow 
\tilde q^\mu = (iq^4, \vec q\,)$ with real $q^4 \in (-\infty,\infty)$,  
the displaced poles will move to the second and forth quadrant. 
Then, after also rotating the momentum 
$k^{\mu} \rightarrow \tilde k^\mu = (ik^4, \vec k)$,  
we obtain the Euclidean vertex functions $f_\beta(\tilde k,P^{(0)})$ 
{}from the Wick rotated BSE:  
\begin{eqnarray} 
        &f_\alpha(\tilde q,P^{(0)})=g^2 \int\frac{d^4k_E}{(2\pi)^4}\,
        &\tilde D_q(-\tilde q-\tilde k-(\eta_q-\eta_D)P^{(0)})\label{WR_BSE}\\
        & &D_q(\eta_qP^{(0)}+\tilde k)\,D_D(\eta_qP^{(0)}-\tilde k)\,
        K_{\alpha \beta}(\tilde q,\tilde k,P^{(0)})\, 
         f_\beta(\tilde k,P^{(0)}),
        \nonumber
\end{eqnarray}
where $d^4k_E = d k^4 d^3\vec k$.

If we are in a kinematic situation where no displaced poles 
occur, i.e. Eq.(\ref{condition_halfWR}) is fulfilled, we may 
obtain the Minkowski space vertex function $f_\alpha(q,P)$ from 
the Euclidean solution through:
\begin{eqnarray}
        &f_\alpha(q,P^{(0)})=g^2 \int_{}^{}\frac{d^4k_E}{(2\pi)^4}\,
        & \tilde D_q(-q-\tilde k-(\eta_q-\eta_D)P^{(0)}) \label{halfWR_BSE}
        \\
        & &\times D_q(\eta_qP^{(0)}+\tilde k)\,D_D(\eta_qP^{(0)}-\tilde k)\,
        K_{\alpha\beta}(q,\tilde k,P^{(0)})\, f_\beta(\tilde k,P^{(0)}).
        \nonumber
\end{eqnarray}
It should be emphasized that for a given Euclidean solution 
$f_\beta(\tilde k,P^{(0)})$,  Eq.(\ref{halfWR_BSE}) 
is not an integral equation but merely an algebraic relation 
between $f_\alpha(q,P^{(0)})$ and $f_\beta(\tilde k,P^{(0)})$.  
If however displaced poles occur, 
i.e. Eq.(\ref{condition_halfWR}) is not fulfilled,
one needs to add contributions from the displaced poles to the RHS of 
Eq.(\ref{halfWR_BSE}). 
This will lead to an inhomogeneous integral equation for the  
function $f_\alpha(q,P^{(0)})$, where the inhomogeneous term
is determined by the Euclidean solution $f_\beta(\tilde k,P^{(0)})$.  

Since the Euclidean solutions $f_\alpha(\tilde q,P^{(0)})$ are functions 
of $\tilde q^2=-q_E^2\equiv-((q_4)^2+|\vec q|^2)$, 
$\tilde q\cdot P^{(0)}=i q_4 M$ for a fixed $M$, it is convenient to introduce 
4-dimensional polar coordinates:
\begin{eqnarray}
        q^4 & = & q_E \cos\alpha_q\,,
        \nonumber \\
        q^i & = & |\vec q\,|\, \hat q^i\,,
        \label{Euclidian_vec} \\
        |\vec q\,| & = & q_E \sin\alpha_q\,.
        \nonumber
\end{eqnarray}
Here $0 < \alpha_q <\pi$ and the 3-dimensional unit vector 
$\hat q^i$ is parameterized by the usual polar and azimuthal angles 
 $\hat q^i=(\sin\theta_q\cos\phi_q,\sin\theta_q\sin\phi_q,\cos\theta_q)$.  
In the following we therefore consider  
$f_\alpha(\tilde q,P^{(0)})$ as a function of $q_E$ and $\cos\alpha_q$.  
Furthermore it is often convenient (and traditional) to factor out   
the coupling constant $g^2$ together with a factor $(4\pi)^2$, and  
define the ``eigenvalue'' $\lambda^{-1}=(g/4\pi)^2$.  
Then the BSE in (\ref{WR_BSE}) is solved as an eigenvalue 
problem for a fixed bound state mass $M$.  

\subsection{$O(4)$ Expansion} \label{O4expansion}

In the following we will define the scalar functions $f_\alpha(q,P)$
for positive energy ($M > 0$) bound states via:
\begin{equation}
        \Gamma (q,P)
                       =\left[ f_1(q,P)+
                        \left(
                              -\frac{P\cdot q}{M^2}
                              +\frac{\myslash{q}}{M}
                \right) f_2(q,P)
        \right]\Lambda^{(+)}(P),
        \label{define_f}
\end{equation}
i.e., we now make a specific choice for the scalar functions $a$ and 
$b$ in Eq.(\ref{BSvertexPara}).  
In the rest frame of this model nucleon, this leads to: 
\begin{equation}
        \Phi^{J^P=1/2^+}_{J_3=S/2}(q,P^{(0)})=
              \Gamma (q,P^{(0)}) u(P^{(0)},S) = 
                \left(\matrix{
                                f_1(q,P^{(0)})
                        \cr
                                \frac{\vec q\cdot\vec\sigma}{M}
                                        f_2(q,P^{(0)})
                        \cr} 
                \right) \chi_S.
        \label{diracrep}
\end{equation}
Here we have explicitly used the Dirac representation.
The Pauli matrices  $\vec\sigma$ act on the two component spinor 
$\chi_S$, where  the spin label $S=\pm 1$ is the 
eigenvalue of $\sigma_3$: $\sigma_3\chi_S=S\,\chi_S$.  
In terms of the O(3) spinor harmonics ${\cal Y}^J_{lm}$
\cite{Edomonds},
\begin{equation}
        {\cal Y}^{1/2}_{0S}(\hat q)=\frac{1}{\sqrt{4\pi}}\chi_S
        \qquad\hbox{and}\qquad 
        {\cal Y}^{1/2}_{1S}(\hat q)=-\hat q\cdot
        \vec \sigma{\cal Y}^{1/2}_{0S}(\hat q),
        \label{O3spnrharmonics}
\end{equation}
we have: 
\begin{equation}
        \Phi^{J^P=1/2^+}_{J_3=S/2}(q,P^{(0)})=\sqrt{4\pi}
                \left(\matrix{
                                f_1(q,P^{(0)})\,{\cal Y}^{1/2}_{0S}(\hat q)
                        \cr
                                -\frac{|\vec q\,|}{M}f_2(q,P^{(0)})
                                        \,{\cal Y}^{1/2}_{1S}(\hat q)
                        \cr}
                \right).
        \label{diracrep2}
\end{equation}
{}From Eq.(\ref{diracrep2}) we observe that 
$f_1$ and $f_2$ correspond to the upper and 
lower components of the model nucleon, respectively.  

After the Wick rotation, as discussed in the previous subsection,
the scalar functions 
$f_\alpha$ become functions of $q_E$ and $\cos\alpha_q$. 
Therefore  we can expand them in terms of Gegenbauer polynomials  $C^1_n(z)$ 
\cite{Betheman}:
\begin{equation}
        f_\alpha(q_E, \cos\alpha_q) 
                = \sum_{n=0}^{\infty}\,i^n\,f^n_\alpha(q_E) 
                        C^1_n(\cos\alpha_q).
        \label{expanf}
\end{equation}
We have introduced the phase $i^n$ to ensure that the coefficient functions
$f^n_\alpha$ are real.  
The integral measure in $O(4)$ polar coordinates is: 
\begin{equation}
        \int\frac{d^4k_E}{(2\pi)^4}=\frac{1}{(4\pi)^2}
        \int_{0}^{\infty}dk_E\,k_E^3\,\frac{2}{\pi}
        \int_{0}^{\pi}d\alpha_k\sin^2\alpha_k\,
        \frac{1}{2\pi}\int d\Omega_{\hat k}.
        \label{measure}
\end{equation}
Multiplying the BSE in Eq.(\ref{WR_BSE}) 
with the Gegenbauer polynomial $C^1_n(\cos\alpha_q)$ 
and integrating over  the hyper-angle $\alpha_q$, 
reduces the BSE to an  integral equation  
for the $O(4)$ radial functions $f^n_\alpha$:
\begin{equation}
        \lambda(M)\,f^n_\alpha(q_E)= \sum_{\beta=1}^{2}\sum_{m=0}^{\infty}\,
                \int_{0}^{\infty}dk_E\, 
                {\cal K}^{n\,m}_{\alpha\,\beta}(q_E,k_E)\; f^m_\beta(k_E).
        \label{fn_eq}
\end{equation}
Here $\lambda(M)$ is the eigenvalue which corresponds to a 
fixed bound state mass $M$.
Furthermore note that the integral kernel 
\begin{eqnarray}
        & {\cal K}^{n\,m}_{\alpha\,\beta}(q_E,k_E) = (-i)^n&i^m\,\frac{2}{\pi}
                \int_{0}^{\pi}d\alpha_q\sin^2\alpha_q\,
                        C^1_n(\cos\alpha_q)\,\frac{2}{\pi}\,
                \int_{0}^{\pi}d\alpha_k\sin^2\alpha_k\,C^1_m(\cos\alpha_k)
        \nonumber\\
        & &\times\frac{1}{2\pi}\int d\Omega_{\hat k}\, k_E^3\,
                \tilde D_q(-\tilde q-\tilde k-(\eta_q-\eta_D)P^{(0)})\, 
                K_{\alpha \beta}(\tilde q,\tilde k,P^{(0)})     
        \label{kernelmatrix}\\
        & &\qquad\times\,D_q(\eta_qP^{(0)}+\tilde k)\,
                D_D(\eta_qP^{(0)}-\tilde k)
        \nonumber
\end{eqnarray}
is real, so that we can restrict ourselves 
to real $O(4)$ radial functions $f^n_\alpha$.

To close this section we shall introduce normalized $O(4)$ radial functions.  
Since the scalar functions $f_1(q, P)$ and $f_2(q, P)$ 
correspond to the upper and lower components of the model nucleon
respectively, one may expect that $f_2(q, P)$ becomes negligible when the 
quark-diquark system forms a weakly bound state.  
Thus one can use the relative magnitude of the  
two scalar functions, $f_{2}(q,P)/f_{1}(q,P)$, as a measure  of 
relativistic contributions to the model nucleon.
To compare the magnitude of the Wick rotated scalar functions 
$f_1(\tilde q, P)$ and $f_2(\tilde q, P)$ 
we introduce normalized $O(4)$ radial functions.  
Recall the $O(4)$ spherical spinor harmonics \cite{Rothe,Ladanyi}:
\begin{equation}
        {\cal Z}_{njlm}(\alpha,\theta,\phi)=
                \left[\frac{2^{2l+1}(n+1)(n-l)!}{\pi(n+l+1)!}\right]^{1/2}\,
                        l!\,(\sin\alpha)^l\,C^{1+l}_{n-l}(\cos\alpha)\, 
                        {\cal Y}^j_{l\,m}(\theta,\phi).
        \label{O4harmocics}
\end{equation}
The integers $n$ and $l$ denote the  $O(4)$ angular momentum and 
the ordinary $O(3)$ orbital angular momentum  respectively.  
The half--integer quantum numbers  
$j$ and $m$ stand for the usual $O(3)$ total angular momentum and the magnetic 
quantum number.  
We rewrite the Wick rotated solution 
$\Phi^{J^P=1/2^+}_{J_3=S/2}(\tilde q,P^{(0)})$ in terms of 
the spinor harmonics 
${\cal Z}_{njlm}(\alpha,\theta,\phi)$ and define the normalized 
$O(4)$ radial functions $F_n(q_E)$ and $G_n(q_E)$ as:
\begin{equation}
        \Phi^{J^P=1/2^+}_{J_3=S/2}(\tilde q,P^{(0)})
        \equiv \sqrt{2}\;\pi 
        \left(\matrix{\sum_{n=0}^{\infty}i^n\,F_n(q_E)\,
                                {\cal Z}_{n\,\frac{1}{2}\,0\,S}(\alpha_q,\hat q)\cr
                \sum_{n=1}^{\infty}i^{n-1}\,G_n(q_E)\,
                        {\cal Z}_{n\,\frac{1}{2}\,1\,S}(\alpha_q,\hat q) \cr}
                 \right).
        \label{O4diracrep}
\end{equation}
The extra factor $\sqrt{2}\;\pi$ is introduced for convenience.  
The normalized $O(4)$ radial functions $F_n$ and $G_n$ are 
then linear combinations of the $f_\alpha^n$:
\begin{eqnarray}
        F_n(q_E) & = & f_1^n(q_E),
        \label{Ffunc} \\
        G_n(q_E) & = & -\frac{q_E}{ 2M }\sqrt{n(n+2)}
                        \left( \frac{f_2^{n-1}(q_E)}{n} 
                                + \frac{f_2^{n+1}(q_E)}{n+2}
                        \right).
        \label{Gfunc}
\end{eqnarray}
Equivalently we can express the Wick rotated scalar functions as:
\begin{eqnarray}
        f_1(q_E,\cos\alpha)&=&
                \sum_{n=0}^{\infty}i^n\,F_n(q_E)\,C^1_n(\cos\alpha),
        \label{f1andF}\\
        f_2(q_E,\cos\alpha)&=&
                -\sum_{n=1}^{\infty}\frac{2M}{q_E}
                        \frac{i^{n-1}}{\sqrt{n(n+2)}}\,
                                G_n(q_E)\,C^2_{n-1}(\cos\alpha).
        \label{f2andG}
\end{eqnarray}

\subsection{Euclidean Solutions}\label{Euclidean}

In this section we present our results for the integral equation 
in (\ref{fn_eq}).  
For simplicity we considered the quark and diquark mass to be equal:
$m_q=m_D=\mb$.  In this case the kernel ${\cal K}_{\alpha \beta}^{n m}$ 
can be evaluated analytically in a simple manner, 
since for $\eta =0$ the denominator 
of the propagator for the exchanged quark does not depend on 
the nucleon momentum $P$.
We fixed the scale of the system by setting the mass $\mb$ to unity.  
The ``mass'' parameter in the form factor was fixed at $\Lambda=2\mb$.  

We solved Eq.(\ref{fn_eq}) as follows.  
First we terminated the infinite sum in Eq.(\ref{expanf}) 
at some fixed value $n_{\rm max}$.  
Then the kernel in Eq.~(\ref{kernelmatrix}) for the  truncated system 
becomes a finite matrix with dimension $(2\times n_{\rm max})^2$. 
Its elements are functions of $q_E$ and $k_E$. 
Next we discretized the Euclidean momenta and performed the integration 
over $k_E$ numerically together with some initially assumed 
radial functions $f^n_\alpha$.  
In this way new radial functions and 
an ``eigenvalue'' $\lambda$ associated with them were generated.  
The value of $\lambda$ was determined by imposing the 
normalization condition on $f^n_\alpha$ such that resultant valence 
quark distribution is properly normalized.  
We then used the generated radial functions as an input and 
repeated the above procedure until the radial functions and 
$\lambda$ converged.  

Note that our normalization differs from the commonly used 
one \cite{Nakanishi_survey},  
since we are going to apply the vertex function only 
to processes with the diquark as a spectator, i.e., we do not consider 
the coupling of the virtual photon 
directly to the diquark.  The ordinary choice of
normalization would not lead to an integer charge for the model
nucleon in the spectator approximation.  We therefore normalize the 
valence quark distribution itself.

Regarding  Eq.(\ref{fn_eq}) as an eigenvalue equation,   
we found the  ``eigenvalue'' $\lambda$ (coupling constant) as a 
function of $M^2$, varying the latter over the range 
$0.85\,\bar m^2 \le M^2 \le 1.99 \,\bar m^2$. 
The eigenvalue $\lambda$ was stable, i.e. independent 
of the number of grid points and the maximum value of $k_E$.
Furthermore, for a weakly bound state, $M> 1.6 \,\bar m$,
the solutions were independent of the choice of the starting functions 
and the iteration converged in typically $10\sim 25$ cycles.  
However, for a strongly bound state, $M < 1.4 \,\bar m$,
we found that the choices of the starting functions were crucial for 
convergence.  A possible reason of this instability for a strongly 
bound system is due to the fact that we did not use the $O(4)$ 
eigenfunctions 
$F_n$ and $G_n$ in numerical calculations but the functions $f_\alpha^n$ 
defined in Eq.~(\ref{expanf}).  
Since strongly bound systems, $M \sim 0$, are approximately  
$O(4)$ symmetric, a truncated set of functions $f_\alpha^n$ may be an  
inappropriate basis for numerical studies of the BSE.  

We found that the eigenvalue $\lambda$ converges quite rapidly 
when $n_{\rm max}$, the upper limit of the 
$O(4)$ angular momentum, is increased.
This stability of our solution with respect to 
$n_{\rm max}$ is independent of $M$. 
We observe that contributions to the eigenvalue $\lambda$  from 
$f_\alpha^n(q_E)$ with $n>4$ are  negligible.  
This dominance of the lowest $O(4)$ radial functions has also been observed 
in the scalar--scalar--ladder model \cite{L+M} and utilized 
as an approximation for solving the BSE in a 
generalized fermion--fermion--ladder approach \cite{Jain+Munczek}.

To compare the magnitude of the two scalar functions 
$f_1(\tilde q, P)$ and $f_2(\tilde q, P)$ we show in Fig.\ref{fig:FandG} 
the normalized $O(4)$ radial functions $F_n$ and $G_n$.
As the dependence of $\lambda$ on $n_{\rm max}$ suggests,   
radial functions with $O(4)$ angular momenta $n>4$ are 
quite small compared to the lower  ones.  
Together with the fast convergence of $\lambda$, 
this observation  justifies  the truncation 
of Eq.(\ref{fn_eq}) at $n=n_{\rm max}$.  
Note that even for very weakly bound systems ($\sim0.5\%$ binding energy) 
the magnitude of the ``lower--component'' $f_2(\tilde q, P)$ remains 
comparable to that of the ``upper--component'' $f_1(\tilde q, P)$.  
This suggests that the spin structure of  relativistic bound states 
is non-trivial,  even for weakly bound systems.  
So-called ``non--relativistic'' approximations, in which one neglects 
the non--leading components of the vertex function ($f_2(\tilde q, P)$ in 
our model), are therefore only valid for extremely weak binding, 
$2\mb \rightarrow M$ only.

\subsection{Analytic Continuation}\label{analyticContinuation}

In the previous section we obtained the quark-diquark vertex 
function in Euclidean space. 
Its application to deep-inelastic scattering, as discussed in 
Sec.\ref{spectatorModel}, demands an analytic continuation to 
Minkowski space.
Here the scalar functions $f_\alpha$, which determine the 
quark-diquark vertex function through Eq.(\ref{BSvertexPara}), 
will depend on the Minkowski space momenta $q^2$ and $P\cdot q$.  

Recall that our Euclidean solution is based on the expansion 
of the scalar functions $f_{\alpha}$ in terms of Gegenbauer polynomials   
in Eq.(\ref{expanf}).  
This expansion was  defined in Sec.\ref{O4expansion} 
for real hyper-angles $\alpha_q$, with  $-1<\cos\alpha_q<1$.
Consequently the infinite sum over the $O(4)$ angular momenta $n$ 
in (\ref{expanf}) is  absolutely convergent for 
pure imaginary energies $q^0$.  
Now we would like to analytically continue $q^0$ to physical, real values.
The Euclidean hyper-angle $\alpha_q$ is defined in Euclidean space such that:  
\begin{equation}
        \cos\alpha_q=\frac{q^0}{i\sqrt{-q^2}}.
        \label{alpha}
\end{equation}
In Minkowski space $\cos\alpha_q$ 
is then purely imaginary ($\cos\alpha_q = -i\,q^0/\sqrt{-q^2}$) 
for space-like $q$,  and real ($\cos\alpha_q = -q^0/\sqrt{q^2}$) 
if $q$ is time-like.  
Note that the angular momentum sum (\ref{expanf}) converges even 
for complex values of $\cos\alpha_q$ as long as  $|\cos\alpha_q|<1$.  
Then an analytic continuation of $f_\alpha$ to Minkowski space
is possible.  
In terms of the Lorentz invariant scalars 
$q^2$ and $P\cdot q$ we obtain:
\begin{equation}
        z =\cos\alpha_q = - 
        \hbox{sgn}(q^2)\,\frac{P\cdot q}{\sqrt{q^2\,M^2}}.
        \label{zMinkowski}
\end{equation}
Then the convergence condition for the sum over the  $O(4)$ angular 
momenta in Eq.(\ref{expanf}) reads: 
\begin{equation}
        (P\cdot q)^2 < M^2 \;|q^2|.
        \label{condition1}
\end{equation}

Even if Eq.({\ref{condition1}) is satisfied 
the radial functions $f_\alpha^n$ 
themselves may contain  singularities which prevent us from performing the 
analytic continuation by numerical methods.  
However, note that the Euclidean solutions for $f_\alpha^n$ are  
regular everywhere on the imaginary $q^0$-axis. 
Consequently the RHS of the  ``half Wick rotated'' equation (\ref{halfWR_BSE})
contains no singularities if the 
displaced-pole-free condition in Eq.(\ref{condition_halfWR}) is met.
Therefore, in Minkowski space, 
the radial functions $f_\alpha$ are regular everywhere 
in the momentum region where the displaced-pole-free condition 
(\ref{condition_halfWR}) and the convergence condition (\ref{condition1}) 
are satisfied.  
Here the analytic continuation to Minkowski space is straightforward. 
Recall the normalized $O(4)$ radial functions $F_n$ 
and $G_n$ from Eqs.(\ref{Ffunc},\ref{Gfunc}), 
which are linear combinations of $f_\alpha^n$.
Writing them as $F_n(q_E)=q_E^n\,\tilde F_n(q_E^2)$ and 
$G_n(q_E)=q_E^n\,\tilde G_n(q_E^2)$,  we find for the 
scalar functions $f_\alpha(q^2,P\cdot q)$ from 
Eqs.(\ref{f1andF},\ref{f2andG}): \hfill
\begin{eqnarray}
        &  & f_1(q^2, P\cdot q)=\sum_{n=0}^{\infty}
                \frac{\tilde F_n(-q^2)}{M^{n}}\,
                \left(\sqrt{q^2\,M^2}\right)^n\,C^1_n(z),
        \label{f1Minkowski} \\
        &  & f_2(q^2, P\cdot q)=-\sum_{n=1}^{\infty}
                \frac{2}{\sqrt{n(n+2)}} 
                \frac{\tilde G_n(-q^2)}{M^{n-2}}\,
                \left(\sqrt{q^2\,M^2}\right)^{n-1}\,
                C^{2}_{n-1}(z).
        \label{f2Minkowski}
\end{eqnarray}
Note that the Gegenbauer polynomials $C^1_n$ ($C^{2}_{n-1}$)
together with the square root factors $(\sqrt{q^2\,M^2})^n$ 
($(\sqrt{q^2\,M^2})^{n-1}$) are $n$-th ($(n-1)$-th) order polynomials of 
$q^2$, $M^2$, and $P\cdot q$ and contain therefore no $\sqrt{q^2\,M^2}$ 
factors.
Since in the kinematic region under consideration, $f_1(q^2, P\cdot q)$ and 
$f_2(q^2, P\cdot q)$ are regular, it is possible to extrapolate 
$\tilde F_n(-q^2)$ and $\tilde G_n(-q^2)$ numerically from space-like $q^2$ to 
time-like $q^2$ as necessary.

Finally we are interested in the quark-diquark vertex function 
as it appears  in the handbag diagram for deep-inelastic scattering.
Therefore we need the functions $f_\alpha$ for  on-shell diquarks only.
The squared relative momentum $q^2$ and the Lorentz scalar $P\cdot q$ 
are then no longer independent but related by:
\begin{equation}
        P\cdot q = -\frac{m_q+m_D}{2 m_D}
                \left[ -q^2 + \left( \frac{m_D}{m_q+m_D} \right) ^2 
                                \left( (m_q+m_D)^2 - M^2 \right) \right].
        \label{on-shellPq}
\end{equation}
Then  $f_1$ and $f_2$ from  Eq.(\ref{f1Minkowski}) and (\ref{f2Minkowski}) 
are functions of the squared relative momentum $q^2$ only.

In Sec.\ref{spectatorModel} the parameterization (\ref{DISpara}) 
for the Dirac matrix structure of the vertex function was more convenient 
to use. 
The corresponding functions $f_{\alpha}^{\rm on}$ 
which enter the nucleon structure function in Eq.(\ref{FofX}) 
are given by : 
\begin{eqnarray}
        & & f^{\rm on}_1(k^2)=f_1(q^2,P\cdot q)
                +\frac{m_D^2-k^2}{2M^2}f_2(q^2,P\cdot q),
        \label{fon1}\\
        & & f^{\rm on}_2(k^2)=\frac{1}{2}f_2(q^2,P\cdot q).
        \label{fon2}
\end{eqnarray}
Here the arguments  $q^2$ and $P\cdot q$, of the scalar functions 
$f_\alpha$ on the RHS should be understood as functions of $k^2$  
through Eq.~(\ref{on-shellPq}), together with the relation:  
\begin{equation}
q^2 = \frac{m_D}{m_q+m_D}\left(k^2 - m_q\left(\frac{M^2}{m_q+m_D} - m_D
\right)\right).
\end{equation}

As already mentioned, the procedure just described yields radial
functions $f_{\alpha}^{\rm on}$ in Minkowski space only in the kinematic
region where the conditions Eqs.(\ref{condition_halfWR},\ref{condition1}) 
are met.
These are fulfilled for weakly bound states $M^2 \lsim (m_D + m_q)^2$ 
at moderate values of $|k^2|$. 
On the other hand, 
the nucleon structure function in (\ref{FofX}) 
at small and moderate values of $x$ is dominated  
by contributions from small quark momenta $|k^2| < m_q^2$.
Consequently, the Minkowski space vertex function obtained in the 
kinematic region specified by the displaced-pole-free condition 
(\ref{condition_halfWR}) and the convergence condition (\ref{condition1})
determines the valence quark distribution of a weakly bound nucleon
at small and moderate $x$. 

In the case of strong binding, $M^2 \ll (m_q + m_D)^2$, or at 
large $x$ the nucleon structure function is dominated by 
contributions from large space-like $k^2$. 
Here the above analytic continuation to Minkowski space is 
not possible and 
the sum over the $O(4)$ angular momenta in Eq.(\ref{expanf}) 
should be evaluated first.
In principle this is possible through 
the Watson-Sommerfeld method \cite{collision_th,Domokos,G+S+G1} 
where the leading power behavior of $f_\alpha(q^2, P\cdot q)$ 
for asymptotic $P\cdot q$ can be deduced by 
solving the BSE at  complex $O(4)$ angular momenta \cite{G+S+G2} , 
or by assuming conformal invariance of the amplitude 
and using  the operator product expansion technique 
as outlined in ref.\cite{G+S+G1}.  

However, the use of the operator product expansion is questionable here, since 
in the quark-diquark model which is being used we have introduced 
a form factor for the quark-diquark coupling and  
our model does not correspond to an asymptotically free theory.  
Existence of the form factor also makes the analysis of 
complex $O(4)$ angular momenta complicated.  
Therefore a simpler approach is used.  
It can be shown from a general analysis  
that BS vertex functions which satisfy a ladder BSE are regular 
for space-like $k^2$, when one of the constituent particles is on 
mass shell.  Furthermore, from the numerical solution studied in the 
previous section, we found that the magnitude of the $O(4)$ partial 
wave contributions to the function $f^{\rm on}_\alpha$ 
decreases reasonably fast for large $O(4)$ angular momenta  $n$,  
except at very large $k^2$.  
We therefore use the expansion formulae 
(\ref{f1Minkowski}) and (\ref{f2Minkowski}) with an upper limit 
on $n \leq n_{\rm max}$ to evaluate $f^{\rm on}_\alpha$ 
defined by Eqs.(\ref{fon1},\ref{fon2}) as an approximation.  
Nevertheless, this application of BS vertex functions to deep-inelastic 
scattering  emphasizes the need to solve Bethe-Salpeter equations 
in Minkowski space from the very beginning, as has been 
done recently for scalar theories without derivative coupling \cite{K+W}. 

\section{Numerical Results}\label{numerical}

In this section we present results for the valence contribution to 
the nucleon structure function, $F_{1}$, from Eq.(\ref{FofX}), 
based on the numerical solutions discussed above.  
First we show in Fig. \ref{fig:fon} the physical, on--shell scalar functions 
$ f^{\rm on}_\alpha$ for a bound state mass 
$M = 1.8\,\bar m$.
The maximal $O(4)$ angular momentum is fixed at  $n_{\rm max}=4$.  
Figure \ref{fig:fon} demonstrates that 
the magnitude of $f^{\rm on}_1$ and 
$f^{\rm on}_2$ is quite similar, even for a weakly bound 
quark-diquark system.  
Furthermore we find that  for weakly bound states 
($M \gsim 1.8 \,\bar m$) 
the dependence of $f^{\rm on}_\alpha$ 
on $n_{\rm max}$ is negligible in the region of moderate,  space-like 
$-k^2 \lsim  5\,\,\bar m^2$.
However  for larger space-like values of $k^2$ the convergence of 
the $O(4)$ expansion in Eq.(\ref{expanf})
decreases for any $M^2$,  and numerical results 
for  fixed $n_{\rm max}$ become less accurate.    

In Fig.~\ref{fig:struct} the structure function, $F_1^{\rm val.}$, is shown 
for various values of $M^2$ using  $n_{\rm max}=4$.
The distributions are  normalized to unity.
One observes that for weakly bound systems ($M = 1.99\,\bar m $) 
the valence quark distribution peaks around $x\sim 1/2$.  
On the other hand, the distribution becomes flat if binding is strong 
($M = 1.2\,\bar m$).  
This behavior  turns out to be mainly of kinematic origin. 
To see this, remember that $F_1^{\rm val.}$ is given by 
an integral (c.f., Eq~ (\ref{FofX}))
over the squared quark momentum, $k^2$, bounded by 
$k^2_{\rm max} = x(M^2 - m_D^2/(1-x))$. 
The latter has a maximum at $x=1-m_D/M$. 
Therefore the peak of the valence distribution for weakly bound 
systems occurs at $x\approx 1/2$ for $m_D = m_q$.
For a more realistic choice $m_D \sim 2\, m_q$ the valence distribution 
would peak at  $x\sim1/3$.
The more strongly the system is bound, the less $k^2_{\rm max}$ varies 
with $x$. This leads to a broad distribution in the case of 
strong binding.
Thus the global shape of $F_1^{\rm val.}$ is determined to a large extent
by relativistic kinematics.

To investigate the role of  the relativistic spin 
structure of the vertex function we discuss the contribution of the 
``relativistic'' component, $f^{\rm on}_2$, to  
the nucleon structure function $F_{1}^{\rm val.}$.  
Figure \ref{fig:weaklybound} shows that   
the contribution from $f^{\rm on}_2$ is negligible for 
a very weakly bound quark-diquark state  $(M = 1.99\,\bar m)$.  
Here the ``non--relativistic'', leading component,  $f^{\rm on}_1$, 
determines the structure function.
However, even for moderate binding the situation is different.
In Fig.\ref{fig:modbound} one observes that 
the contribution from the ``relativistic'' component is quite significant 
for $M = 1.8\,\bar m$. 
Nevertheless, the characteristic $x$ dependence, i.e. the  
peak of the structure function at $x \approx 1/2$, is still due to the  
``non--relativistic'' component.


\section{Summary}\label{summary}

The aim of this work was to outline a scheme whereby 
structure functions can be obtained from a relativistic 
description of a model nucleon as a quark-diquark bound state.
For this purpose we solved the BSE for the 
nucleon starting from a simple quark-diquark Lagrangian.
{}From the Euclidean solutions of the BSE we extracted 
the physical quark-diquark vertex functions. 
These were applied to the spectator model for DIS, and the 
valence quark contribution to the structure function 
$F_1$ was calculated.

Although the quark-diquark Lagrangian used here is certainly not 
realistic, and the corresponding BSE was solved by applying 
several simplifications,
some interesting and useful observations were made.
We found that the spin structure of the nucleon, 
seen as a relativistic quark-diquark bound state, is non-trivial,
except in the case of very weak binding.  Correspondingly, the valence 
quark contribution to the structure function is governed by the 
``non--relativistic'' component of the nucleon vertex function only 
for a very weakly bound state. 
Furthermore, we observed that the shape of the unpolarized 
valence quark distribution is mainly determined by 
relativistic kinematics and does not depend on details 
of the quark-diquark dynamics.   
However at large quark momenta 
difficulties in the analytic continuation of 
the Euclidean solution for the vertex function 
to Minkowski space emphasize the need to treat 
Bethe-Salpeter equations in Minkowski space from 
the very beginning.

\section*{Acknowledgments}

This work was supported in part by the the Australian Research Council, 
BMBF and the Scientific Research grant \#1491 of Japan Ministry of Education, 
Science and Culture.

\section*{\Large\bf Figure Captions}

\noindent
Fig. \ref{fig:diquarkSpectatorForT}:  
The diquark spectator contribution to the virtual forward 
Compton amplitude in the Bjorken limit. 
\vspace{1.0cm}

\noindent
Fig. \ref{fig:ladderBSE}:  
The Bethe-Salpeter equation for a quark-diquark system in the ladder 
approximation.
\vspace{1.0cm}

\noindent
Fig. \ref{fig:FandG}:  
The normalized $O(4)$ radial functions $F_n$ and $G_n$ 
{}from (\ref{O4diracrep}) for $M=1.8\,\bar m$ as 
functions of the Euclidean momentum $q_E$. 
\vspace{1.0cm}

\noindent
Fig. \ref{fig:fon}:  
The on--shell scalar functions $f^{\rm on}_1$ (solid) and 
$-f^{\rm on}_2$ (dashed) as a function of the quark 
momentum, $k^2$, for $M=1.8\,\bar m$ and $n_{\rm max}=4$.
\vspace{1.0cm}

\noindent
Fig. \ref{fig:struct}:  
The valence quark distribution, $F_1^{\rm val.}$, from Eq.~(\ref{FofX}) 
for different  binding for the model proton. 
The solid, dashed, and dot--dashed lines show the results for 
weak ($M=1.99 \, \bar m$), 
moderate ($M=1.8 \, \bar m$), 
and strong ($M=1.2 \, \bar m$) binding 
respectively.
\vspace{1.0cm}

\noindent
Fig. \ref{fig:weaklybound}:
Contribution of the ``relativistic  component",
$f^{\rm on}_2$, to the structure function $F_1^{\rm val.}$ from 
Eq.~(\ref{FofX}) for a  weakly bound model proton
($M=1.99\,\bar m$).  
The solid line denotes the total valence distribution, $F_1^{\rm val.}$. 
The dashed line shows the contributions to $F_1^{\rm val.}$, 
which are proportional to $f^{\rm on}_2$.
\vspace{1.0cm}

\noindent
Fig. \ref{fig:modbound}:  
As in Fig.\ref{fig:weaklybound}, but for moderate binding 
for the model proton ($M=1.8\,\bar m$).
\vspace{1.0cm}

\newpage
\begin{figure}[t]
        \centering{\epsfig{figure=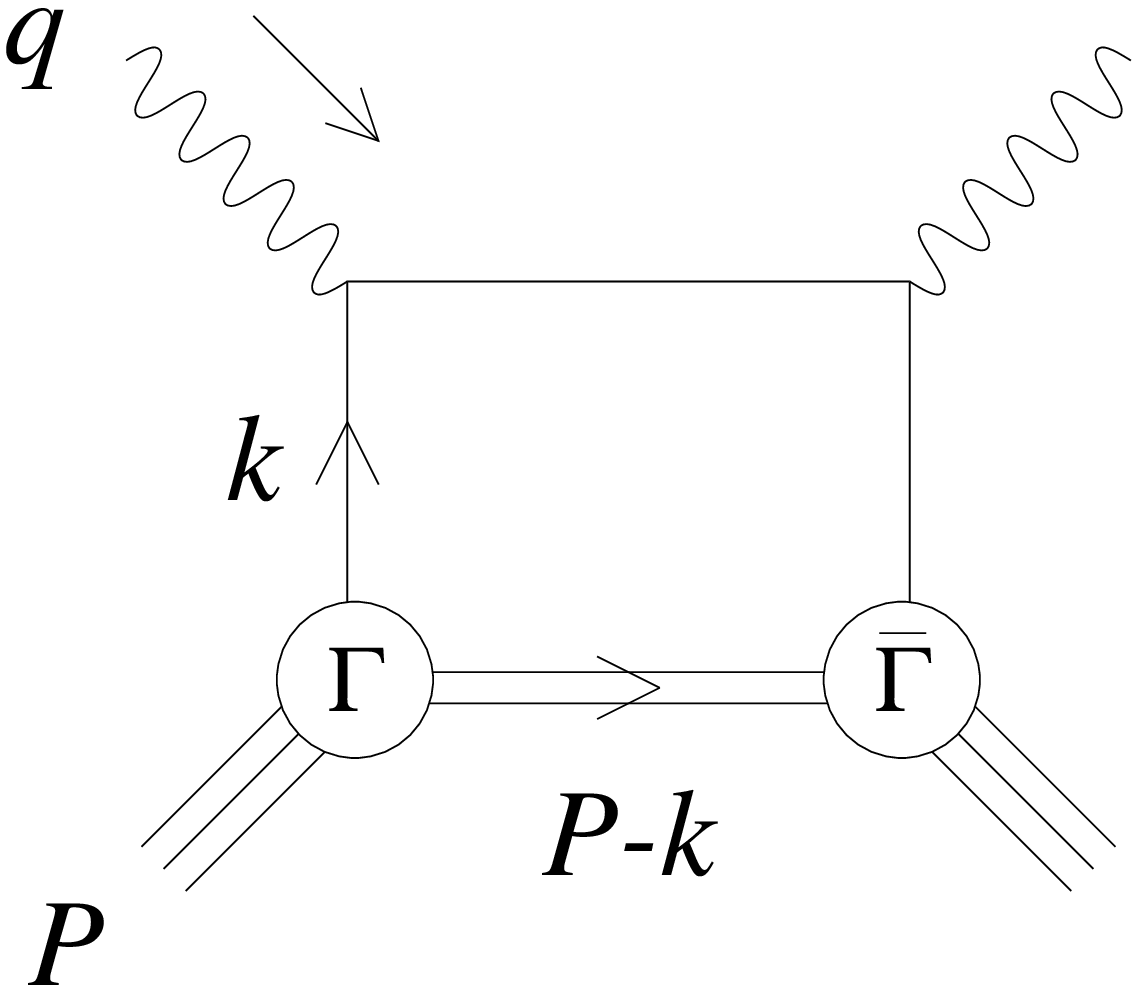,height=10.0cm} }
        \vspace{1.0cm}
        \caption{}
        \label{fig:diquarkSpectatorForT}
\end{figure}

\newpage
\begin{figure}[t]
        \centering{\epsfig{figure=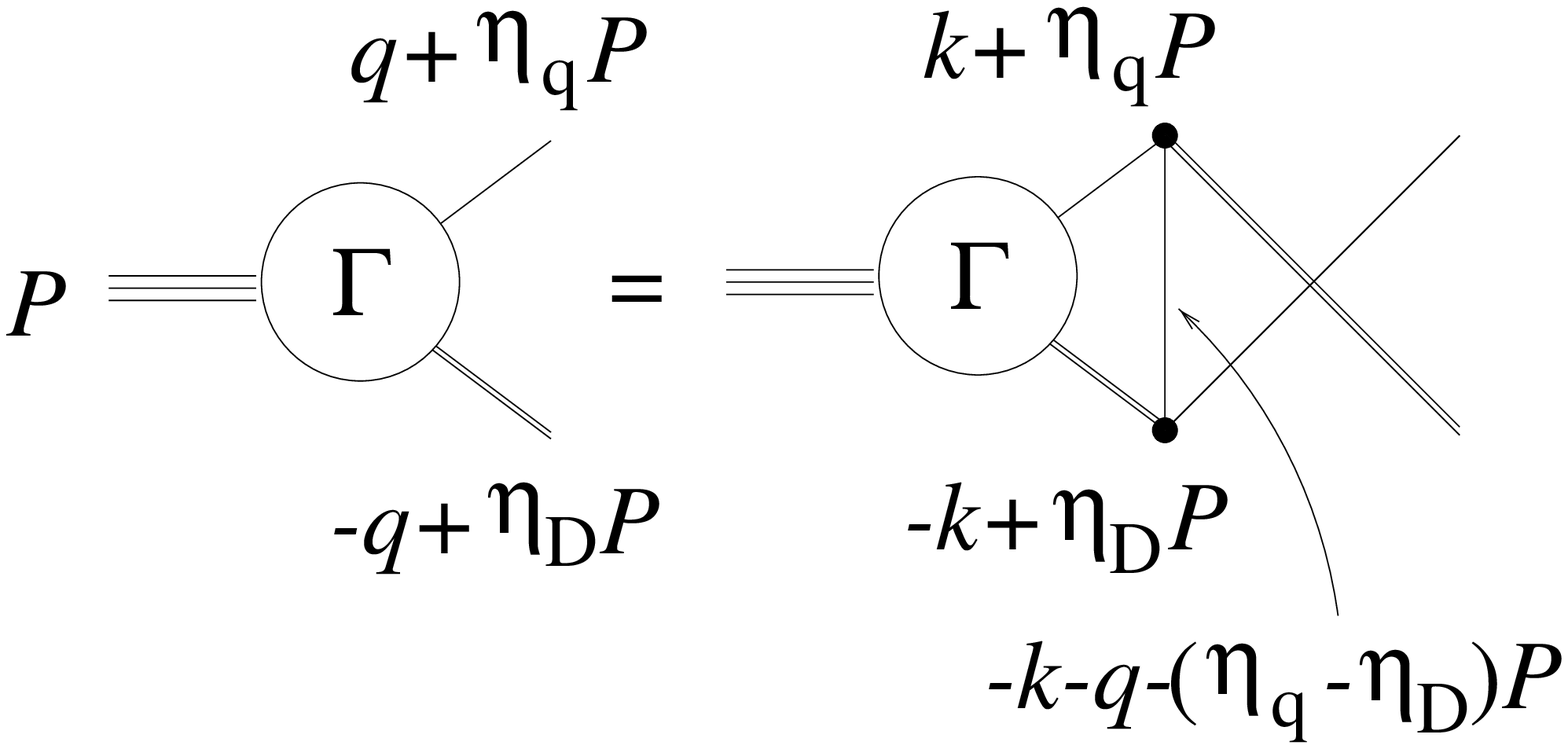,height=7.0cm} }
        \vspace{1.0cm}
        \caption{}
        \label{fig:ladderBSE}
\end{figure}

\newpage
\begin{figure}[t]
        \centering{\epsfig{figure=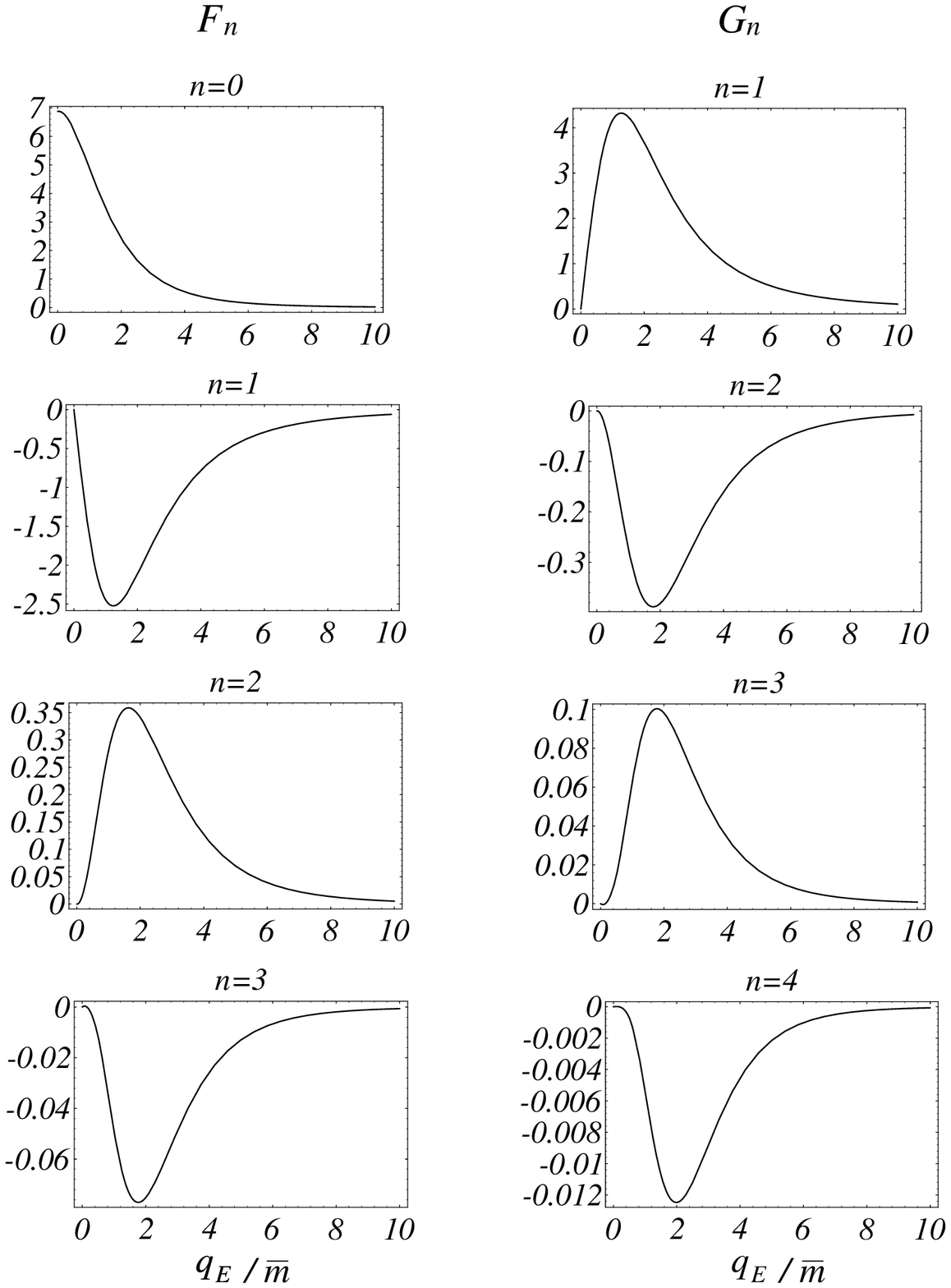,height=17.5cm} }
        \vspace{1.0cm}
        \caption{}
        \label{fig:FandG}
\end{figure}

\newpage
\begin{figure}[t]
        \centering{\epsfig{figure=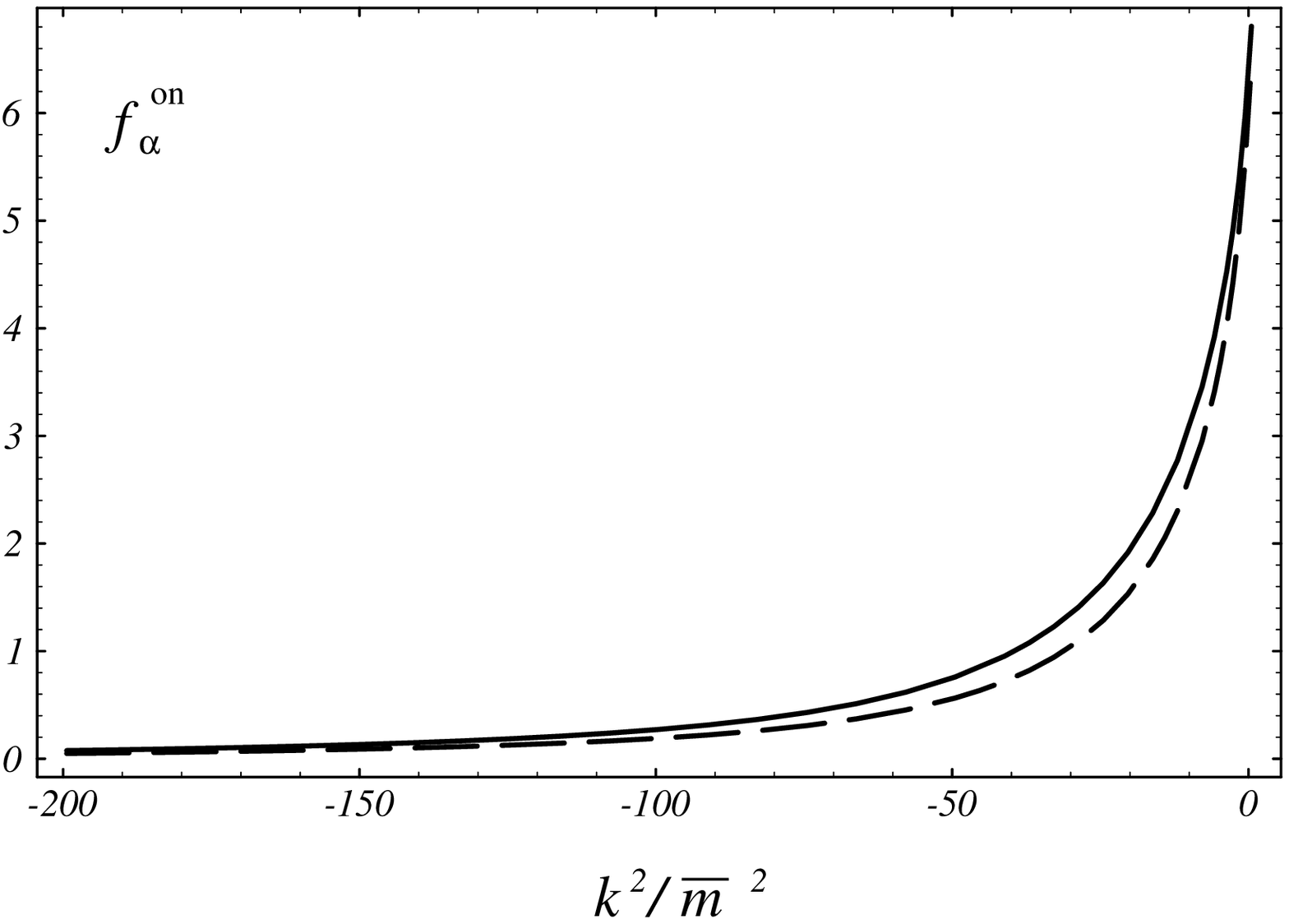,height=17.5cm} }
        \vspace{1.0cm}
        \caption{}
        \label{fig:fon}
\end{figure}

\newpage
\begin{figure}[t]
        \centering{\epsfig{figure=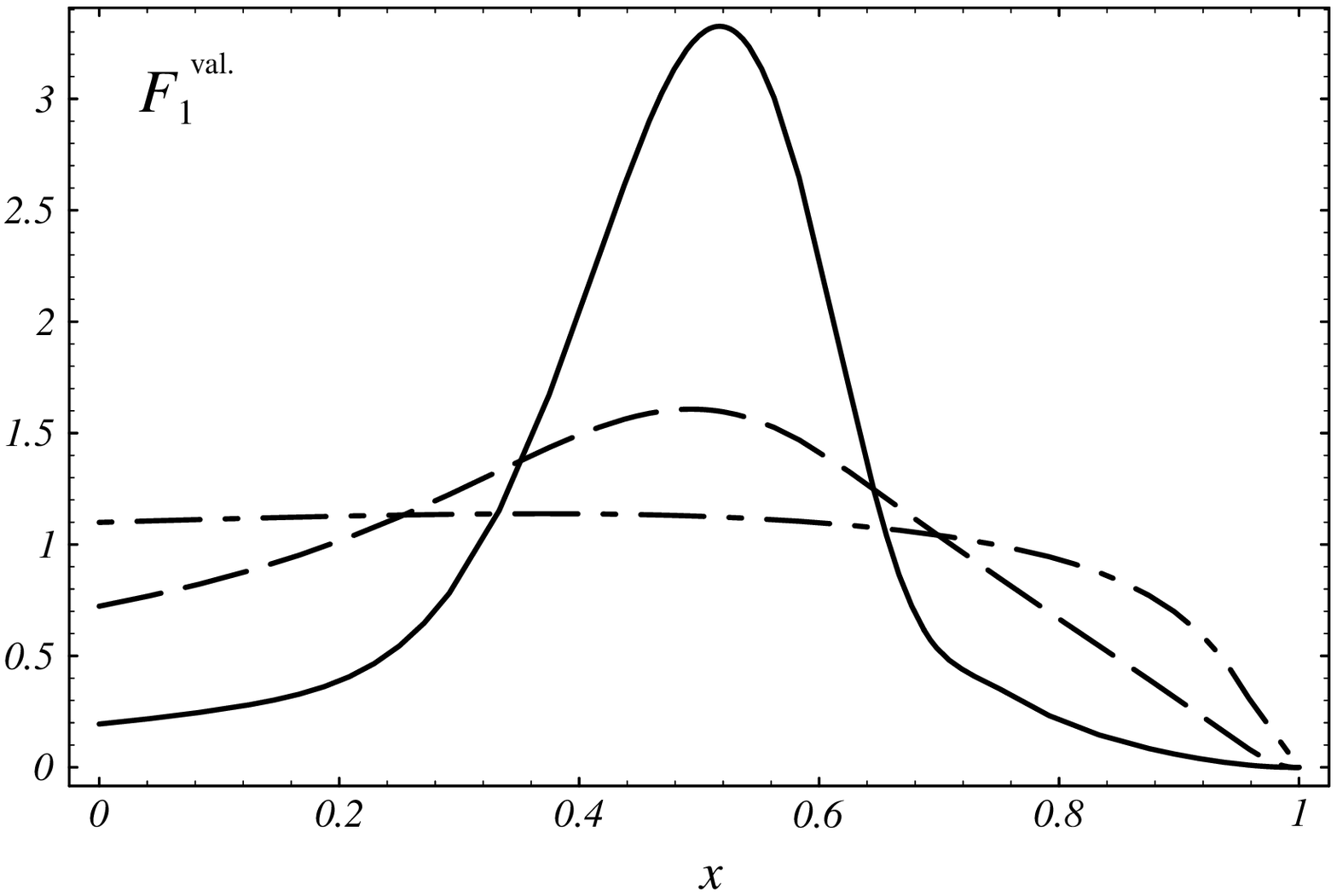,height=17.5cm} }
        \vspace{1.0cm}
        \caption{}
        \label{fig:struct}
\end{figure}

\newpage
\begin{figure}[t]
        \centering{\epsfig{figure=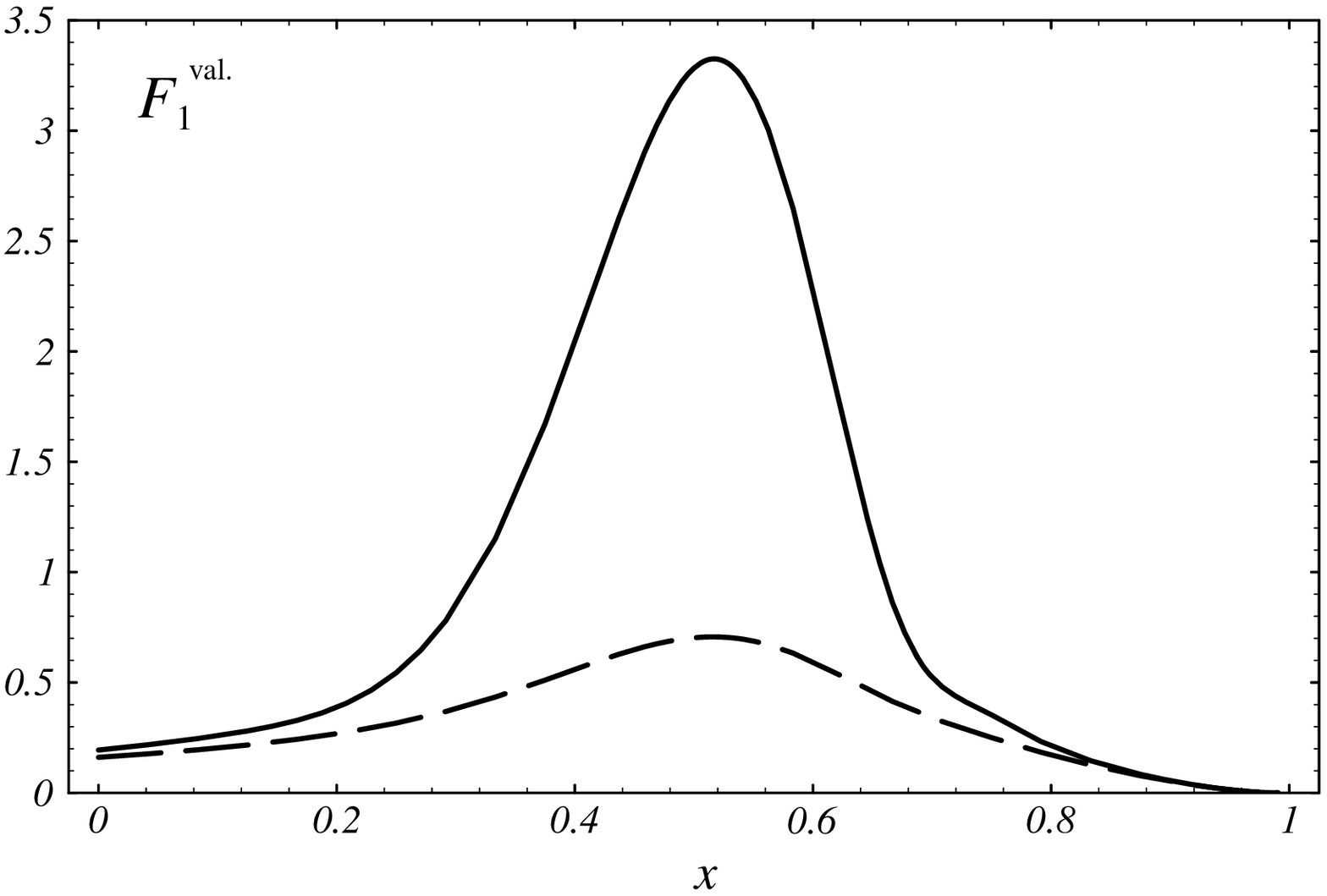,height=17.5cm} }
        \vspace{1.0cm}
        \caption{}
        \label{fig:weaklybound}
\end{figure}

\newpage
\begin{figure}[t]
        \centering{\epsfig{figure=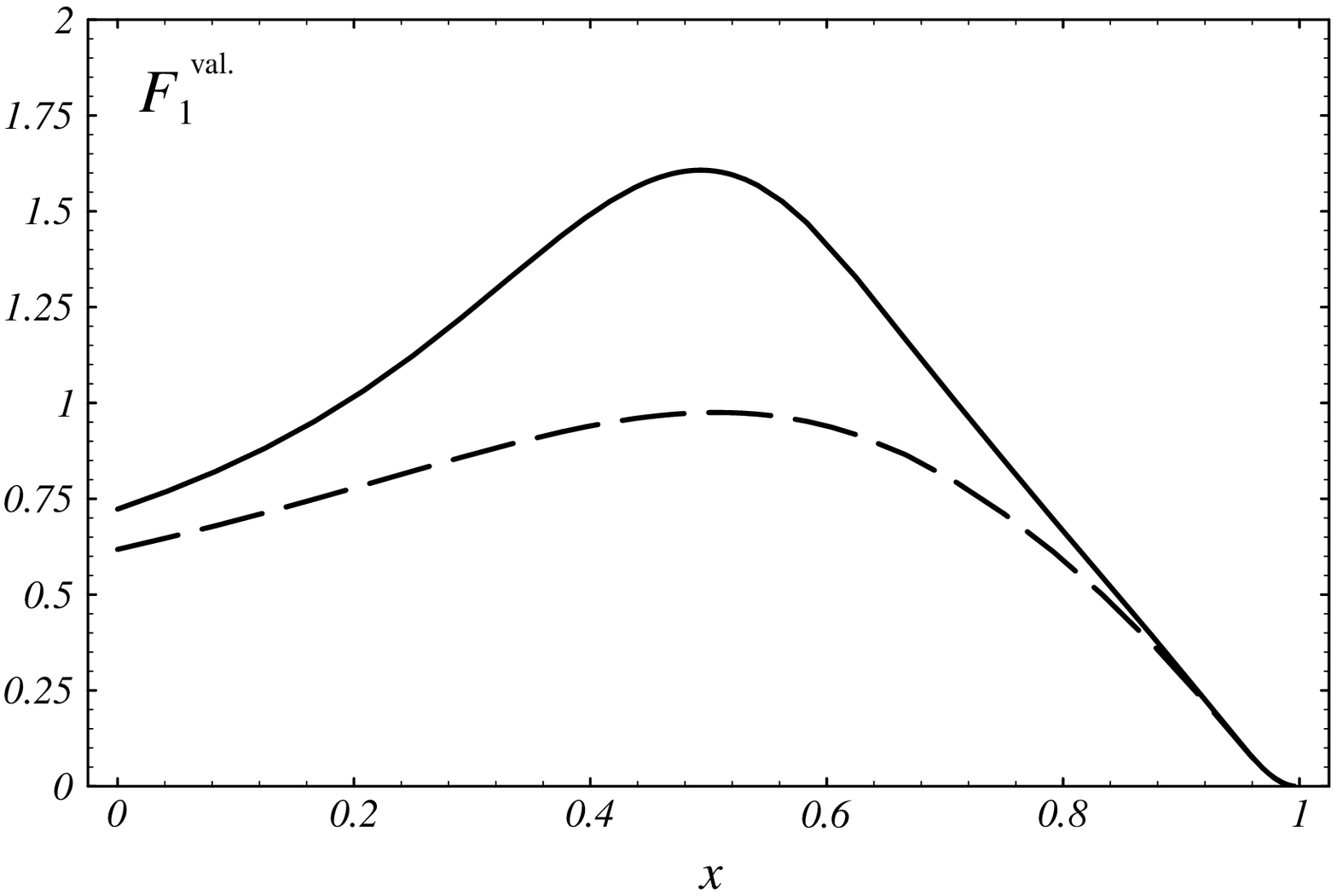,height=17.5cm} }
        \vspace{1.0cm}
        \caption{}
        \label{fig:modbound}
\end{figure}

\end{document}